\documentclass{moriond}

\usepackage[utf8]{inputenc}
\usepackage[T1]{fontenc}
\usepackage{lmodern}
\usepackage{textcomp}
\usepackage{microtype}

\usepackage[english]{babel}
\usepackage[autostyle, english = american]{csquotes}
\MakeOuterQuote{"}

\usepackage{amsmath}
\usepackage{mathtools}
\usepackage{upgreek}
\usepackage{units}
\usepackage{slashed}
\usepackage{physics}

\DeclareMathOperator{\diag}{diag}
\DeclareMathOperator{\BR}{BR}

\usepackage{booktabs}
\usepackage{graphicx}

\newcommand{\Photo}{}

\title{Perspectives to find heavy neutrinos with NA62 \footnote{Contribution to the proceedings of the 53rd Rencontres de Moriond on Electroweak Interactions and Unified Theories (2018) based on Ref.~\cite{Drewes:2018gkc}.}}

\author{Marco Drewes$^{1,2}$ \footnote{Speaker},
Jan Hajer$^{1,3}$,
Juraj Klaric$^{4,2}$,
Gaia Lanfranchi$^5$
}

\address{\footnotesize\strut\\
$^1$Centre for Cosmology, Particle Physics and Phenomenology, Universit\'e catholique de Louvain,\\Louvain-la-Neuve B-1348, Belgium\\
$^2$Excellence Cluster Universe, Boltzmannstr. 2, D-85748, Garching, Germany\\
$^3$Institute for Advanced Study, The Hong Kong University of Science and Technology,\\Clear Water Bay, Kowloon, Hong Kong S.A.R., China\\
$^4$Physik Department T70, Technische Universit\"at M\"unchen,\\James Franck Stra\ss e 1, D-85748 Garching, Germany\\
$^5$Laboratori Nazionali INFN di Frascati, Frascati, Italy
}

\begin{document}

\maketitle

\abstracts{The sensitivity of beam dump experiments to heavy neutrinos depends on the relative size of their mixings with the lepton flavours in the Standard Model.
We study the impact of present neutrino oscillation data on these mixing angles in the minimal type I seesaw model.
We find that current data significantly constrains the allowed heavy neutrino flavour mixing patterns.
Based on this, we discuss the implications for the sensitivity of the NA62 experiment to heavy neutrinos when operated in the beam dump mode.
We find that NA62 is currently the most sensitive experiment in the world for heavy neutrino masses between that of the kaon and the $D$-mesons.
The sensitivity can vary by almost two orders of magnitude if the heavy neutrinos exclusively couple to the tau flavour, but depends only comparably weakly on the flavour mixing pattern within the parameter range preferred by light neutrino oscillation data.}

\section{Introduction}

\subsection{Motivation}

All elementary fermions in the Standard Model (SM) of particle physics with the exception of neutrinos are known to exist with both chiralities, left handed and right handed.
If right handed neutrinos exist, they could possibly explain several open puzzles in particle physics and cosmology, cf.~e.g.~\cite{Drewes:2013gca} for an overview.
Most importantly, they can explain the light neutrino flavour oscillations via the \emph{type I seesaw mechanism}~\cite{Minkowski:1977sc}. 
A key prediction of this mechanism is the existence of heavy neutral leptons (HNL) $N_i$ with masses $M_i$ and weak interactions with the SM flavours $a=e,\mu,\tau$ that are suppressed by small mixing angles $\theta_{a i}$.
Further motivation comes from cosmology.
The interactions of right handed neutrinos generally violate $CP$ and can potentially generate a matter-antimatter asymmetry in the primordial plasma, which can be converted into a net baryon number by weak sphalerons~\cite{Kuzmin:1985mm}.
This process known as \emph{leptogenesis}~\cite{Fukugita:1986hr} provides a possible explanation for the \emph{baryon asymmetry of the universe} (BAU), which is believed to be the origin of baryonic matter in the present day universe (cf.~e.g.~\cite{Canetti:2012zc} for a discussion).
It can either occur during the freeze-out and decay of the $\nu_R$~\cite{Fukugita:1986hr} ("freeze-out scenario") or during their production~\cite{Akhmedov:1998qx,Asaka:2005pn,Hambye:2016sby} ("freeze-in scenario").
The freeze-in scenario is e.g.~realised in the \emph{Neutrino Minimal Standard Model} ($\nu$MSM)~\cite{Asaka:2005pn}. 
It is particularly interesting from a phenomenological viewpoint because it is feasible for masses $M_i$ as low as \unit[10]{MeV}~\cite{Canetti:2012kh}, which are well within reach of present day experiments~\cite{Chun:2017spz}.
The NA62 experiment~\cite{NA62:2017rwk} can probe part of this mass range.

\subsection{The seesaw model}

The most general renormalisable Lagrangian that can be constructed from SM fields and $n$ flavours of right handed neutrinos $\nu_{R i}$ reads
\begin{equation}
    \mathcal{L}
  = \mathcal{L}_\text{SM} + i \overline{\nu_{R i}}\slashed\partial\nu_{R i}
  - \frac{1}{2} \left( \overline{\nu_{R i}^c}(M_M)_{ij}\nu_{R j} + \overline{\nu_{R i}}(M_M^\dagger)_{ij}\nu_{R j}^c \right)
  - F_{a i}\overline{\ell_{L a}}\varepsilon\phi \nu_{R i}
  - F_{a i}^*\overline{\nu_{R i}}\phi^\dagger \varepsilon^\dagger \ell_{L a}
\ . \label{eq:Lagrangian}
\end{equation}
Here $\varepsilon$ is the antisymmetric SU(2) tensor and we have suppressed SU(2) indices.
$M_M$ is a Majorana mass matrix for the right handed neutrinos, and the $F_{a i}$ are Yukawa couplings between the $\nu_{R i}$ and the SM leptons $\ell_a$.%
\footnote{Throughout this document we use four component spinor notation. The chiral spinors $\nu_R$ and $\ell_L$ have only two non-zero components ($P_R\nu_R=\nu_R$ and $P_L\ell_L=\ell_L$). As a result, no explicit chiral projectors are required in the weak interaction term \eqref{WeakWW}.}
After electroweak symmetry breaking the Higgs field obtains an expectation value
$v = 174$ GeV, which generates the Dirac mass term $\overline{\nu_{L}} m_D\nu_{R}$ with $m_D=vF$ from the term $F\overline{\ell_L}\varepsilon\phi \nu_{ R }$.
The three light and $n$ heavy mass eigenstates after electroweak symmetry breaking can be expressed in terms of the Majorana spinors
\begin{align}
    \upnu_i &
  = \left[
  V_{\nu}^{\dagger}\nu_L-U_{\nu}^{\dagger}\theta \nu_R^c+V_{\nu}^T\nu_L^c-U_{\nu}^T\theta^{\ast} \nu_R
  \right]_i
\ ,
  & N_i &
  = \left[
  V_N^\dagger\nu_R+\Theta^T \nu_L^c + V_N^T\nu_R^c+\Theta^{\dagger}\nu_L
  \right]_i
\ , \label{HeavyMassEigenstates}
\end{align}
respectively.
Here $V_\nu = (1 - \frac{1}{2}\theta\theta^\dagger ) U_\nu$, where $U_\nu$ is the matrix which diagonalises the light neutrino mass matrix
\begin{equation}
    m_\nu
  = - m_D M_M^{-1} m_D^T
  = - \theta M_M \theta^T \quad \text{with} \quad \theta=m_D M_M^{-1} = v F M_M^{-1},
\label{seesaw}
\end{equation}
as $U_\nu^\dagger m_\nu U_\nu^* = \diag(m_1,m_2,m_3)$,
while $V_N = (1 - \frac{1}{2} \theta^T \theta^*) U_N$, where $U_N$ is the equivalent matrix that diagonalises the heavy neutrino mass matrix $M_N = M + \frac{1}{2} (\theta^\dagger \theta M + M^T \theta^T \theta^{*})$ as $U_N^T M_N U_N = \diag(M_1,M_2,\ldots,M_n)$
after electroweak symmetry breaking.
The matrix $\Theta = U_N^*\theta$ mixes the "active" and "sterile" neutrinos $\nu_L$ and $\nu_R$, leading to a $\theta$-suppressed weak interaction of the heavy mass eigenstates $N_i$,
\begin{align}
    \mathcal L
  \supset&
  - \frac{g}{\sqrt{2}}\overline{N}_i \Theta^\dagger_{i \alpha}\gamma^\mu e_{L a} W^+_\mu
 - \frac{g}{2\cos\theta_W}\overline{N_i} \Theta^\dagger_{i a}\gamma^\mu \nu_{L a} Z_\mu
 - \frac{g}{\sqrt{2}}\frac{M_i}{m_W}\Theta_{a i} h \overline{\nu_{L a}}N_i
+ \text{h.c.}
\ . \label{WeakWW}
\end{align}
The first two terms are the couplings of the $N_i$ to the weak currents, and the last term is the Yukawa coupling to the physical Higgs field $h$ in the unitary gauge,
for which we have used the relation $m_W = \frac{1}{2}v g$ involving the weak gauge coupling constant $g$.
This Yukawa term is not relevant for NA62 because the event rate for processes mediated by virtual Higgs bosons at NA62 is suppressed by the small Yukawa couplings of the first generation fermions involved in the kinematically accessible final states.
Due to the interactions~\eqref{WeakWW} the $N_i$ can appear in all processes that involve ordinary neutrinos if this is kinematically allowed, but with amplitudes suppressed by the angles~$\Theta_{a i}$.
It is convenient to express event rates in terms of the quantities
\begin{align}
U_{a i}^2 &= \abs{\Theta_{a i}}^2
\ ,
 &  U_a^2
 &= \sum_i U_{a i}^2
\ ,
 &  U_i^2
 &= \sum_a U_{a i}^2
\ ,
 &  U^2
 &= \sum_i U_i^2
\end{align}
because the HNL production and decay rates are proportional to combinations of the $U_{a i}^2$~\cite{Gorbunov:2007ak}.
The seesaw relation~\eqref{seesaw} suggests that $U_i^2 \sim \sqrt{\Delta m_\text{atm}^2 + m_\text{lightest}^2} / M_i < \unit[10^{-10}]{GeV}/M_i$, which would clearly imply unobservably small branching ratios in experiments.
If, however, the Lagrangian~\eqref{eq:Lagrangian} approximately respects the $B-L$ symmetry of the SM~\cite{Shaposhnikov:2006nn}, then much larger $U^2$ are possible because the symmetry leads to systematic cancellations in $m_\nu m_\nu^\dagger$, and the light neutrino masses must be proportional to small parameters that measure the amount of $B-L$ violation.
This implies that heavy neutrinos with mixings $U_i^2$ that are much larger than the above estimate must be of the pseudo-Dirac type, i.e., must be organised as in pairs $N_i$ and $N_j$ with $M_i \simeq M_j$ and $Y_{ia} \simeq i Y_{ja}$, so that $U_{a i}^2 \simeq U_{a j}^2 \simeq U_a^2/2$.

\subsection{The NA62 experiment}

The NA62 experiment~\cite{NA62:2017rwk} is a fixed target experiment located at CERN's SPS beam.
With a nominal beam intensity of $3\times 10^{12}$ protons per pulse and pulses of \unit[4.8]{s} it can collect up to $3\times 10^{18}$ protons on target (POT) per year.
NA62 can be operated in two different modes, the \emph{kaon mode} or \emph{target mode} on one hand and the \emph{dump mode} on the other.
The mode of operation can be changed in \unit[15]{min}.

\paragraph{Kaon mode.} The primary goal of the NA62 experiment for which it is currently taking data is to measure the branching ratio (BR) of the $K^+ \to \pi^+ \nu \overline{\nu}$ decay with a precision of at least \unit[10]{\%}. In order to achieve this goal the experiment needs to collect about $10^{13}$ kaon decays in its normal operation mode, the kaon mode.
In this mode the primary \unit[400]{GeV} proton beam impinges on a \unit[400]{mm} long cylindrical beryllium target with a diameter of \unit[2]{mm}, producing a secondary positively charged hadron beam with a momentum of \unit[75]{GeV}.
The secondary beam reaches the \unit[120]{m} long evacuated decay volume which has a diameter of \unit[2]{m} about \unit[100]{m} downstream of the target.
The kaons, which make up about \unit[6]{\%} of the hadron beam, are identified and timestamped by a N\textsubscript{2} filled Cherenkov counter located along the beam line.
Charged particles from kaon decays inside the decay volume are detected by
a ring-imaging Cherenkov (RICH) counter filled with Neon which separates $\pi$, $\mu$ and $e$ for momenta up to \unit[40]{GeV}.
Their time of flight is measured both by the RICH and by the scintillator hodoscopes placed downstream of the RICH. The forward region is covered by an electromagnetic calorimeter.
The hadronic calorimeter provides further separation between $\pi$ and $\mu$ based on hadronic energy, while a fast scintillator array identifies muons with sub-nanosecond time resolution.
In this mode, NA62 can search for $N_i$ with masses below the kaon mass by looking for a peak in the spectrum of charged leptons produced along with the $N_i$~\cite{CortinaGil:2017mqf}.

\paragraph{Dump mode.}

In the \emph{dump mode}, the target is pulled up and the primary proton beam
is send directly onto the Cu-Fe based collimators, which act as a hadron stopper (or \emph{dump}) located \unit[20]{m} downstream of the target.
The various hadrons produced in the collision with the dump can decay into HNLs in the detector volume with branching ratios determined in~\cite{Gorbunov:2007ak}.
Those HNLs travel downstream into the detector volume, where they can further decay into SM particles.
The signal signature is therefore a vertex of two (or more) tracks appearing in the middle of the fiducial volume and nothing else.
In the present work~\cite{Drewes:2018gkc} we estimate the sensitivity of NA62 to HNLs decaying into at least two-track final states.
As a basis for our computations we use a dataset of $10^{18}$ POT, which will be collected during Run 3 (2021--2023).
From those, roughly $2 \times 10^{15}$ $D$-mesons and $\sim 10^{11}$ $b$-hadrons are produced.
The sensitivity to HNLs crucially depends on their "flavour mixing pattern", i.e.  the relative size of their couplings to the individual SM flavours, which can be characterised by the ratios $U_{ai}^2/U_i^2$.

\section{HNL flavour mixing pattern}

\begin{figure}
\includegraphics[width = 0.49\textwidth]{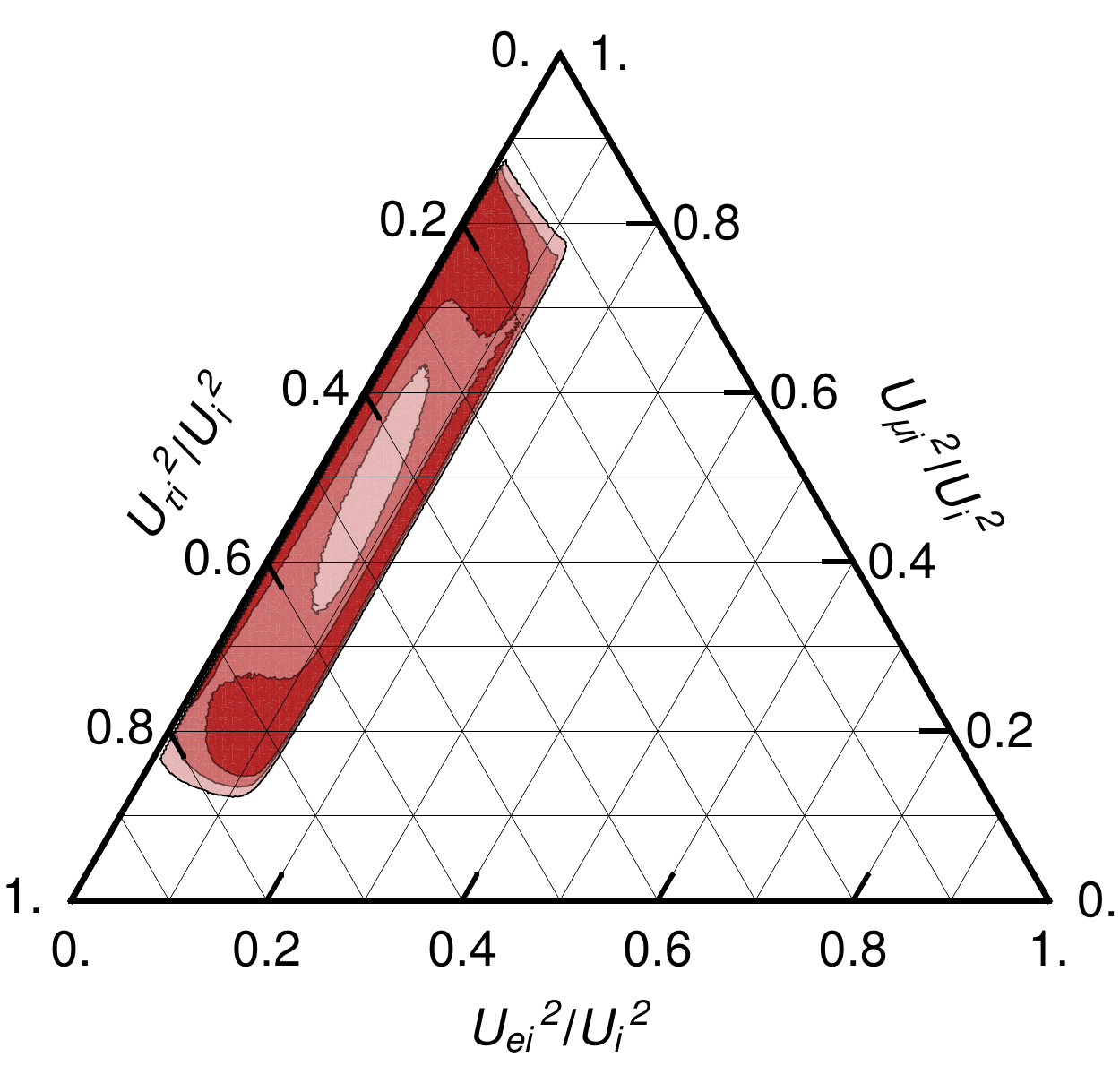}\hfill
\includegraphics[width = 0.49\textwidth]{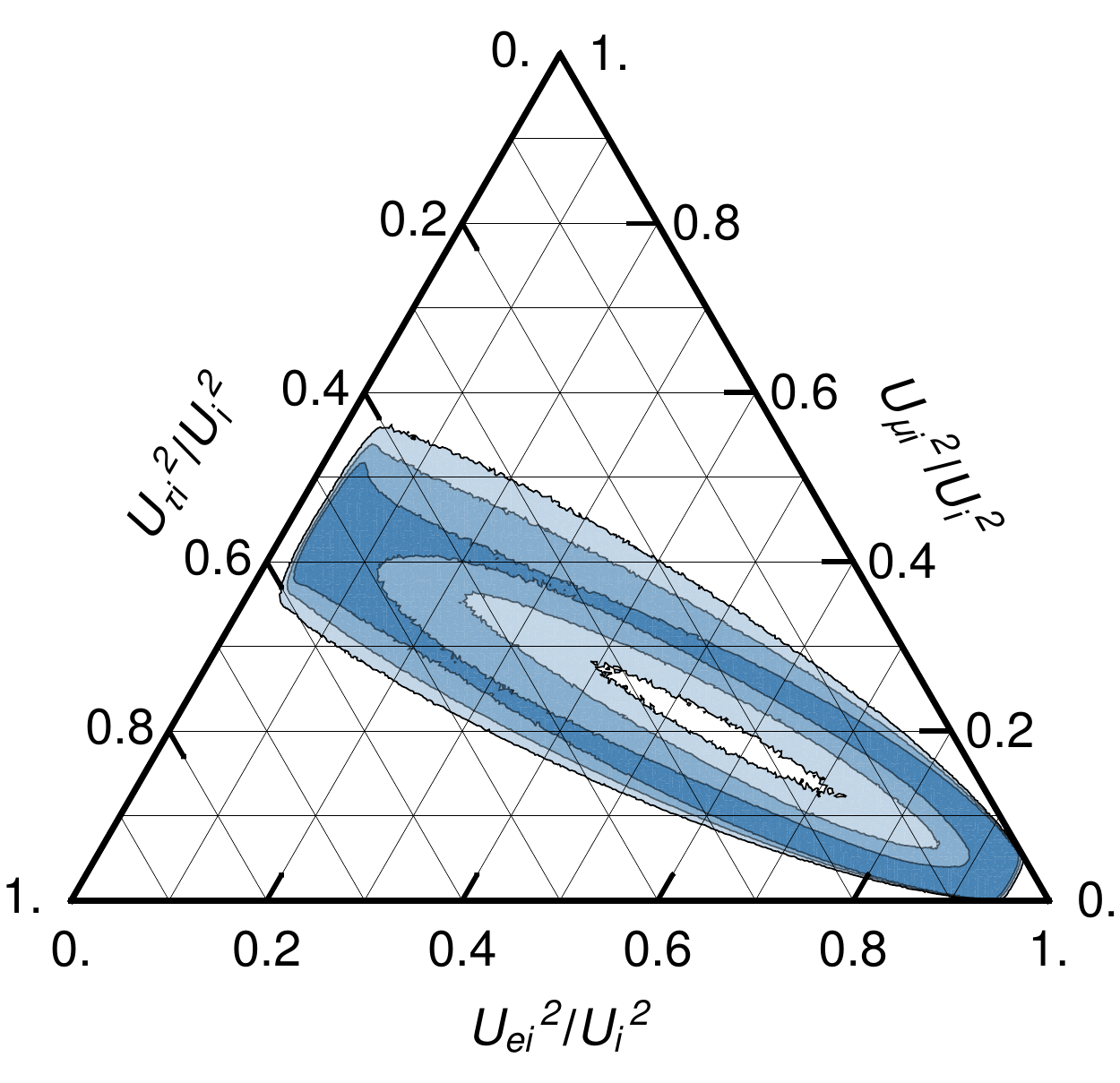}
\caption{The different shades indicate the 1$\sigma$ (darkest), 2$\sigma$ and 3$\sigma$ (lightest) probability contours for the ratios $U_{ai}^2/U_i^2$ for $n = 2$  that can be obtained from present neutrino oscillation data from the NuFIT~3.2 global fit with normal ordering (left) or inverted ordering (right) of the light neutrino masses. 
\label{fig:triangles}
}
\end{figure}

\begin{table}
\centering
\begin{tabular}{cc@{ : }c@{ : }cccc}
 \toprule
 & \multicolumn{3}{c}{Ratio}
 & \multicolumn{3}{c}{Percent of $U_i^2$}
 \\ \cmidrule(r){2-4}
 \cmidrule(l){5-7}
 & $U_{e i}^2$
 & $U_{\mu i}^2$
 & $U_{\tau i}^2$
 & $U_{e i}^2$
 & $U_{\mu i}^2$
 & $U_{\tau i}^2$
 \\ \midrule
 A)
 & 1
 & 160
 & 27.8
 & 0.530
 & 84.7
 & 14.7
 \\ B)
 & 1
 & 1.71
 & 5.62
 & 12.0
 & 20.5
 & 67.5
 \\ C)
 & 1
 & 10.5
 & 15.9
 & 3.65
 & 38.3
 & 58.0
 \\ D)
 & 1
 & 0
 & 0
 & 100
 & 0
 & 0
 \\ E)
 & 0
 & 1
 & 0
 & 0
 & 100
 & 0
 \\ F)
 & 0
 & 0
 & 1
 & 0
 & 0
 & 100
 \\ \bottomrule
\end{tabular}
\caption{Benchmark scenarios used in this analysis.\label{tab:benchmark scenarios}
}
\end{table}

The ratios $U_{ai}^2/U_i^2$ depend on the light neutrino oscillation parameters in $U_\nu$~\cite{Shaposhnikov:2008pf}. 
We consider the cases $n=2$ and $n=3$, which correspond to the minimal number of $\nu_{R i}$ needed to explain the two observed light neutrino mass differences and a scenario in which $n$ equals the number  of fermion generations in the SM, respectively.
In the minimal model with $n=2$ statistically significant posterior probabilities for the ratios $U_{ai}^2/U_i^2$ can be derived 
from present light neutrino oscillation data~\cite{Esteban:2016qun}, cf.~Fig.~\ref{fig:triangles}.
Details of the analysis are given in Ref.~\cite{Drewes:2018gkc}.
The choice $n=2$ effectively also describes the $\nu$MSM because the constraints on the mass and mixings of the third HNL from the requirement that it is a viable Dark Matter candidate (cf.~\cite{Adhikari:2016bei}) imply that it can practically be neglected in the present context.
The scenarios A)--D) in table~\ref{tab:benchmark scenarios} are motivated by the allowed regions in Fig.~\ref{fig:triangles}.

For $n=3$ in principle all values of $U_{ai}^2/U_i^2$ can be made consistent with light neutrino oscillation data.
However, for a hierarchical spectrum of light neutrino masses values outside the allowed regions in Fig.~\ref{fig:triangles} can only be achieved with tuning in the model parameters. Hence, the mass of the lightest neutrino, which is expected to be measured from cosmological data in the foreseeable future, determines how precisely one can predict the allowed range of the $U_{ai}^2/U_i^2$.

\begin{figure}
\includegraphics[width = 0.49\textwidth]{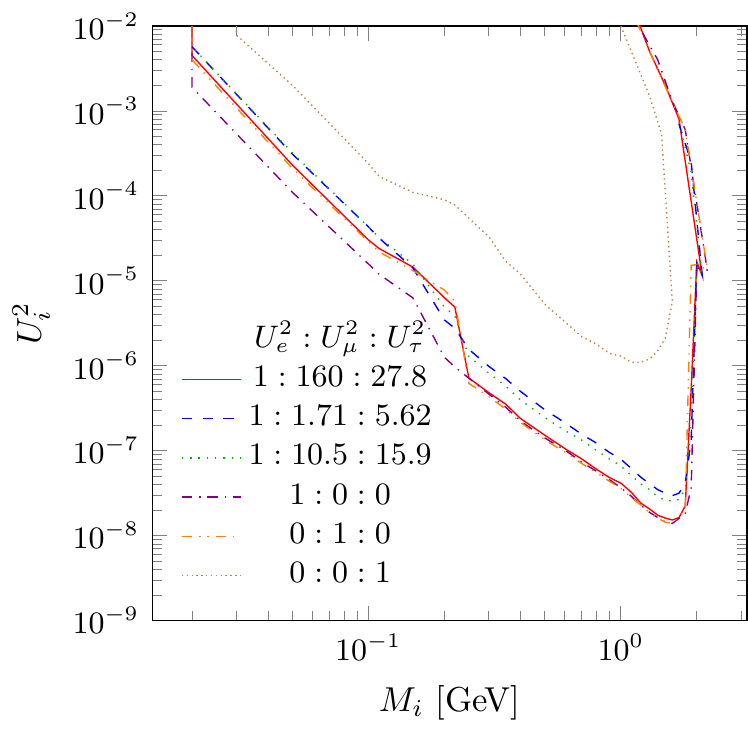}\hfill
\includegraphics[width = 0.49\textwidth]{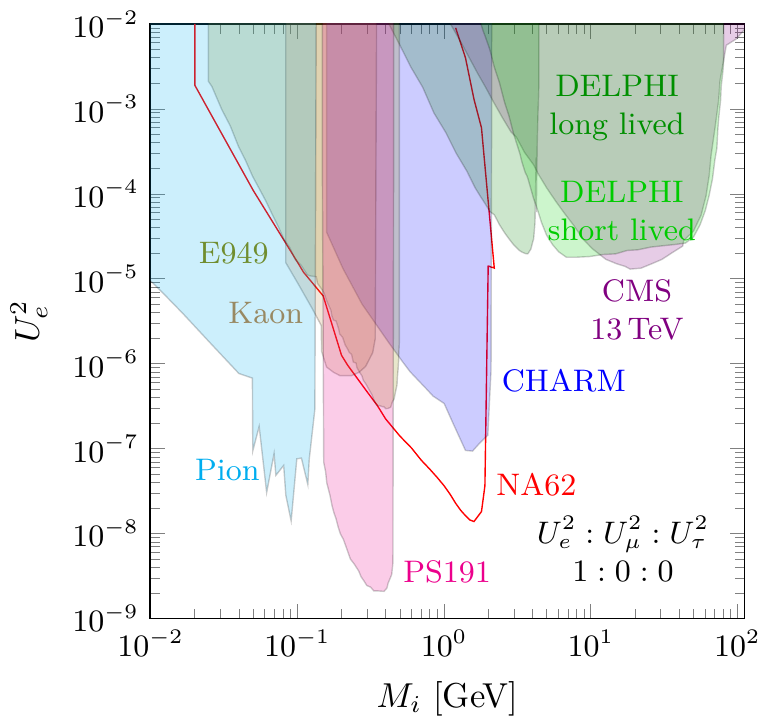}\\
\includegraphics[width = 0.49\textwidth]{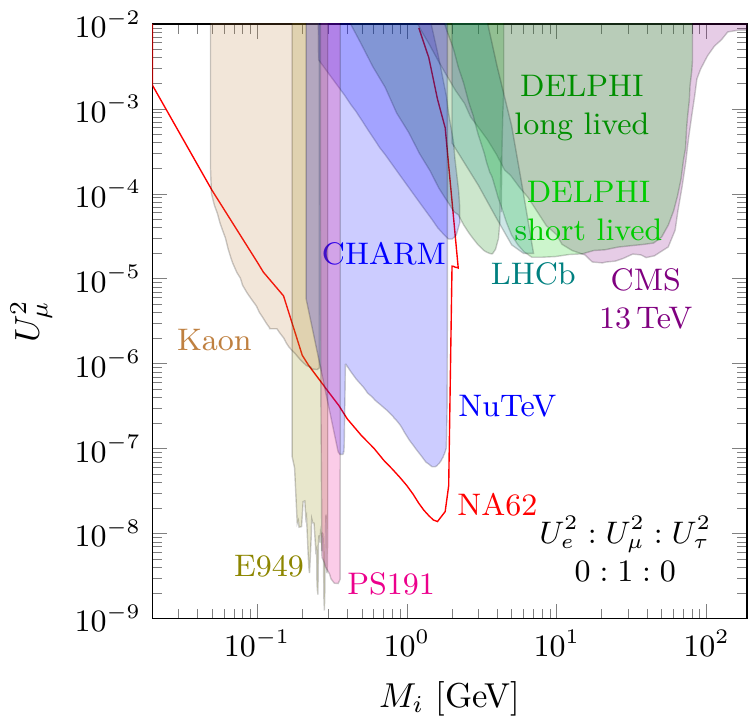}\hfill
\includegraphics[width = 0.49\textwidth]{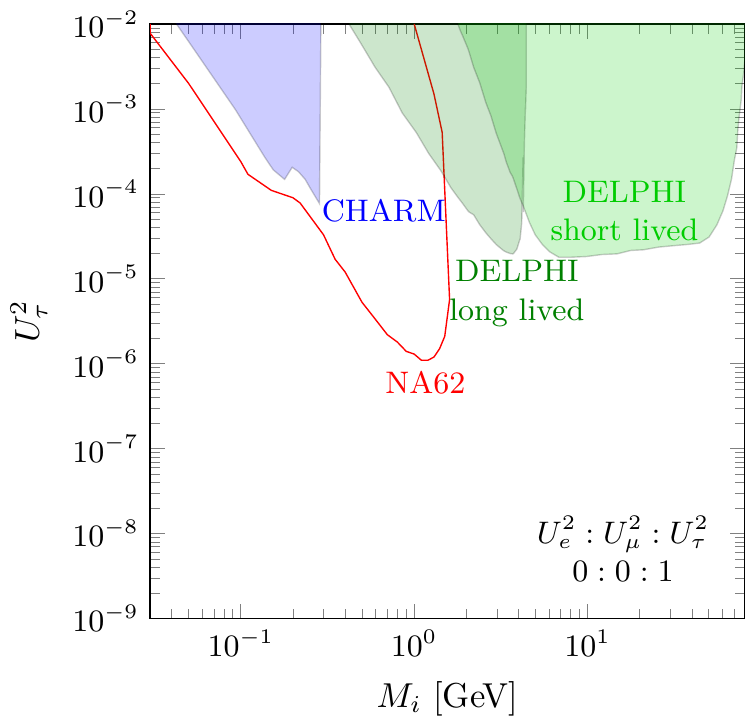}\\
\caption{%
\emph{Upper left panel}: Sensitivity of the NA62 experiment in the scenarios A)--F) in table~\ref{tab:benchmark scenarios}.
The region above the curves marks the expected exclusion regions for 2.3 events for each scenario.
\emph{Remaining panels}: Comparison of the NA62 sensitivity to the published exclusion regions from past experiments (shaded areas) for $U_e^2$ (upper right), $U_\mu^2$ (lower left) and $U_\tau^2$ (lower right).
The regions covered by past experiments have been obtained from Refs.~\protect\cite{Sirunyan:2018mtv} (CMS),~\protect\cite{Abreu:1996pa} (DELPHI),~\protect\cite{Antusch:2017hhu} (LHCb~\protect\cite{Aaij:2016xmb}),~\protect\cite{Atre:2009rg} (CHARM~\protect\cite{Bergsma:1985is}, CHARM II~\protect\cite{Vilain:1994vg}, NuTeV~\protect\cite{Vaitaitis:1999wq} and combined kaon and pion decay bounds), E949~\protect\cite{Artamonov:2014urb} and~\protect\cite{Ruchayskiy:2011aa} (PS191~\protect\cite{Bernardi:1987ek}).
We only show the strongest constraints, a more complete discussion of existing bounds can e.g.~be found in Ref.~\protect\cite{Drewes:2015iva}.
}
\label{fig:ExtremeScenarios}
\end{figure}

\section{NA62 sensitivity}

The computation of the NA62 sensitivity for $U^2_{ia}$ ($a=e,\:\mu,\:\tau$) in different scenarios in Ref.~\cite{Drewes:2018gkc} was performed using a toy Monte Carlo in which all the kinematics of the $N_i$ production and decay processes have been implemented.
The geometrical acceptance for the decay products has been evaluated using the geometry of the experiment as described in Ref.~\cite{NA62:2017rwk}, but we assume the background to be fully negligible.
A detailed discussion of the background in NA62 in the beam dump mode can be found in Ref.~\cite{Lanfranchi:2017wzl}.
In the sensitivity computation we assumed $10^{18}$ POT.
The $c$- and $b$-hadrons can originate from primary protons or from all secondary products of the hadronic shower in the dump (in particular protons, neutrons, and pions).
We studied the composition of the shower and the kinematics of the produced $c$- and $b$-hadrons by simulating the \unit[400]{GeV} proton beam on a thick ($\sim 11 \lambda_I$) high-$Z$ target with \texttt{Pythia}~6.4~\cite{Sjostrand:2006za}.
Given the ratio $\sim 10^4$ between $c$- and $b$-hadron decays, the $N_i$ production via charm decays is the dominant process up to the $D$-meson masses.
The contribution from $b$-hadrons decays at the NA62 intensity is almost negligible.
The $N_i$ decay to SM particles via the same $\theta$-suppressed weak interactions that are responsible for their production.
The main decay channels of $N_i$ in the mass range below the $D$-mesons are
\begin{equation}
    N_i
  \to
3 \nu ,\:
\pi^0 \nu ,\:
\pi^{\pm} \ell^{\mp} ,\:
\rho^0 \nu ,\:
\rho^{\pm} l ,\:
\ell^+ \ell^- \nu
\label{eqn:decays}
\end{equation}
where $\ell = e,\:\mu,\:\tau$.
The NA62 detector is able to reconstruct all the final states with two charged tracks.
The number of events reconstructed in the NA62 detector is given by
\begin{equation}
    N_\text{obs}
  = \sum_{\mathllap{I=\text{prod}}\text{uction m}\mathrlap{\text{odes}}} n_{N,I}
    \sum_{f,\:f'\mathrlap{=e,\:\mu,\:\tau,\:\pi,\:K}}
    \BR \left(N_i \to f^+ f^{\prime-} X \right)
    \mathcal A_i\left( f^+ f^{\prime-} X,\: M_i, U^2_{e,\:\mu,\:\tau}\right)
    \varepsilon \left(f^+ f^{\prime-} X,\: M_i \right)
\ ,
\label{eq:Nobs}
\end{equation}
where $n_{N,I}$ is the number of $N_i$ produced in the a given production process $I$.
$\mathcal A_I (f^+ f^{\prime-} X,\:\allowbreak M_i,\:\allowbreak U^2_{e,\:\mu,\:\tau})$ is the geometrical acceptance for a $N_i$ of a given mass $M_i$ and coupling $U^2_{e,\:\mu,\:\tau}$ that is produced in that process $I$ and decays into a final state with two charged tracks ($f^+ f^{\prime-}$) and other decay products $X$ (e.g.~photons and neutrinos).
The expected sensitivity for the scenarios A)--F) defined in table~\ref{tab:benchmark scenarios} is shown in Fig.~\ref{fig:ExtremeScenarios}.

\section{Discussion and conclusions}

Our results show that NA62, when operated in the dump mode, is currently the world's most sensitive experiment for HNLs with masses between the kaon and $D$-meson mass, cf. Fig.~\ref{fig:ExtremeScenarios}.
The sensitivity to the HNL coupling to $\tau$ exceeds that of past experiments by several orders of magnitude,  
while the expected improvement for $e$ and $\mu$ is expected to be roughly half an order of magnitude.
NA62 is likely to remain as the world's most powerful tool to search for HNLs in this mass range until one of the dedicated experiments that have been proposed (e.g. SHiP \cite{Anelli:2015pba}, MATHUSLA \cite{Chou:2016lxi}, FASER \cite{Feng:2017uoz}, CODEX-b \cite{Gligorov:2017nwh}) is built.

The NA62 sensitivity in principle depends on the HNL's "flavour mixing pattern", i.e., the relative size of their coupling to individual SM flavours. It primarily depends on the ratio $U_{\tau i}^2/(U_{ei}^2 + U_{\mu i}^2)$; for $M_i$ above the dimuon threshold it is practically independent of the ratio $U_{ei}^2/U_{\mu i}^2$.
In the minimal model with $n=2$ heavy neutrinos, which effectively also describes the $\nu$MSM, large values of $U_{\tau i}^2/(U_{ei}^2 + U_{\mu i}^2)$ are ruled out by neutrino oscillation data, cf.~Fig.~\ref{fig:triangles}, so that the sensitivity is almost independent of the flavour mixing pattern. 

\subsection*{Acknowledgements}

This research was supported by the Collaborative Research Center SFB1258 of the Deutsche Forschungsgemeinschaft and by the DFG cluster of excellence "Origin and Structure of the Universe" (\url{universe-cluster.de}).
Jan Hajer was supported by the Research Grants Council of the Hong Kong S.A.R.\ under the Collaborative Research Fund (CRF) Grant \textnumero\ HUKST4/CRF/13G.

\end{document}